\documentclass[9pt,sigconf,screen,nonacm]{acmart}
\AtBeginDocument{%
  }

\setcopyright{none}

\usepackage{bm}

\usepackage{subcaption}

\usepackage{quantikz}
\usepackage{makecell}

\theoremstyle{plain}
\newtheorem{fact}{Fact}

\usepackage[capitalize]{cleveref}
\crefname{fact}{Fact}{Facts}
\crefname{algocf}{Algorithm}{Algorithms}

\usepackage{enumitem}
\setlist[itemize]{leftmargin=6mm}

\usepackage[ruled,vlined]{algorithm2e}

\DeclareMathOperator{\stab}{Stab}
\newcommand{\cnot}{\mathit{CNOT}}

\newcommand{\FF}{\mathbb{F}}
\newcommand{\PP}{\mathbb{P}}
\newcommand{\cD}{\mathcal{D}}
\newcommand{\cE}{\mathcal{E}}
\newcommand{\cO}{\mathcal{O}}

\newcommand{\overto}[1]{\overset{#1}{\to}}
\newcommand{\red}[1]{{\color{red!75!black}#1}}

\usepackage{tikz}
\usetikzlibrary{patterns}

\usepackage{pgfplots}
\pgfplotsset{
    compat=1.5,
    grid=major,
}
\usetikzlibrary{plotmarks}

\def\COMPILETIKZ{0}

\usetikzlibrary{external}

\ifthenelse{\COMPILETIKZ=1}{
\tikzexternalize[prefix=figures/]%
\tikzexternaldisable
}{}

\newcommand{\tikzinput}[1]{
\ifthenelse{\COMPILETIKZ=1}
{%
{
\tikzexternalenable\tikzsetnextfilename{#1_tikz}\input{./figures/#1}%
}
}%
{\includegraphics{./figures/#1_tikz.pdf}}%
}

\newcommand{\symphase}{\texttt{SymPhase.jl}}

\usepackage{titlesec}
\newcommand{\periodafter}[1]{#1.}
\titleformat{\subsubsection}[runin]{\normalfont\normalsize\bfseries\boldmath}{\thesubsubsection}{.5em}{\periodafter}
\titlespacing{\subsubsection}{0pt}{*.5}{*1}

\begin{document}

\title{SymPhase: Phase Symbolization for Fast Simulation of Stabilizer Circuits}


\author[W. Fang]{Wang Fang}\orcid{0000-0001-7628-1185}
\affiliation{
  \department{State Key Laboratory of Computer Science}
  \institution{Institute of Software, Chinese Academy of Sciences}
  \city{Beijing}
  \country{China}
}
\email{fangw@ios.ac.cn}
\affiliation{
  \institution{University of Chinese Academy of Sciences}
  \city{Beijing}
  \country{China}
}

\author[M. Ying]{Mingsheng Ying}\orcid{0000-0003-4847-702X}
\affiliation{
  \department{State Key Laboratory of Computer Science}
  \institution{Institute of Software, Chinese Academy of Sciences}
  \city{Beijing}
  \country{China}
}
\email{yingms@ios.ac.cn}
\affiliation{
  \department{Department of Computer Science and Technology}
  \institution{Tsinghua University}
  \city{Beijing}
  \country{China}
}


\begin{abstract}
  This paper proposes an efficient stabilizer circuit simulation algorithm that only traverses the circuit forward once.
We introduce phase symbolization into stabilizer generators, which allows possible Pauli faults in the circuit to be accumulated explicitly as symbolic expressions in the phases of stabilizer generators.
This way, the measurement outcomes are also symbolic expressions, and we can sample them by substituting the symbolic variables with concrete values, without traversing the circuit repeatedly. 
We show how to integrate symbolic phases into the stabilizer tableau and maintain them efficiently using bit-vector encoding.
A new data layout of the stabilizer tableau in memory is proposed, which improves the performance of our algorithm (and other stabilizer simulation algorithms based on the stabilizer tableau).
We implement our algorithm and data layout in a Julia package named \symphase{}, and compare it with Stim, the state-of-the-art simulator, on several benchmarks.
We show that \symphase{} has superior performance in terms of sampling time, which is crucial for generating a large number of samples for further analysis.
\end{abstract}



\maketitle

\section{Introduction}\label{sec:intro}
With the rapid development of quantum hardware, designing and building large-scale fault-tolerant quantum computer architecture has become an urgent task~\cite{2023arXiv230807915B,2023arXiv230808648X,Acharya2022SuppressingQE,abobeih2022fault,zhao2022realization}.
It relies on quantum error correction (QEC) protocols, for which the implementation relies on stabilizer circuits~\cite{gottesman1997stabilizer}.
Due to the complexity and unintuitive nature of quantum systems, it is essential to have efficient methods for simulating stabilizer circuits on classical computers, as this can help us design and test circuits and protocols before deploying them on quantum hardware, like classical EDA tools.

Fortunately, stabilizer circuits are a special class of quantum circuits that can be simulated in polynomial time on classical computers~\cite{gottesman1998heisenberg}.
There are several efficient stabilizer circuit simulators available~\cite{gidney2021stim,aaronson2004improved,anders2006fast,krastanov}, but they are still not sufficient for analyzing fault-tolerant gadgets.
A typical example is that we need to repeatedly sample the faults that occur inside the circuit of a gadget and count the measurement outcomes of the circuit under these fault samples to evaluate the performance of the gadget.
Existing simulators can generate a single sample very fast, but the number of samples can be in the millions when the circuit is large, making the simulation very slow. 
The state-of-the-art stabilizer circuit simulator, Stim, also mentioned that generating samples of QEC circuits remains the bottleneck in analysis~\cite{gidney2021stim}.

\begin{figure}[t]
    \hspace{-3mm}\scalebox{.8}{\begin{tikzpicture}
    \node (a) {\setlength{\arraycolsep}{0pt} \begin{quantikz}[row sep=0.2cm]
        \lstick{\ket{0}} &[-1mm] \gate{H} &[-3mm] \ctrl{1} &[-3mm] \qw &[-3mm] \qw &[9mm] \gate{Z^{\red{s_1}}} &[9mm] \qw &[-3mm] \qw &[-3mm] \ctrl{1} &[-3mm] \gate{H} &[-1mm] \meter{}\rstick{$m_1$} \\
        \lstick{\ket{0}} & \qw & \targ{} & \ctrl{1} &\qw & \gate{X^{\red{s_2}}} & \qw & \ctrl{1} & \targ{} & \qw &\meter{}\rstick{$m_2$} \\
        \lstick{\ket{0}} & \qw & \qw & \targ{} & \ctrl{1} & \gate{X^{\red{s_3}}} & \ctrl{1} & \targ{} & \qw & \qw & \meter{}\rstick{$m_3$}\\
        \lstick{\ket{0}}\slice[style={blue!70},label style={pos=1, anchor=north}]{$\begin{array}{c}\ket{\psi_0}={}\\ \begin{array}{lcccc}(-1)^0&Z&I&I&I\\ (-1)^0&I&Z&I&I\\ (-1)^0&I&I&Z&I\\ (-1)^0&I&I&I&Z \end{array} \end{array}$} & \qw & \qw & \qw & \targ{}\slice[style={blue!70},label style={pos=1, anchor=north}]{$\begin{array}{c}\ket{\psi_1}={}\\ \begin{array}{lcccc} (-1)^0&X&X&X&X\\ (-1)^0&Z&Z&I&I\\ (-1)^0&I&Z&Z&I\\ (-1)^0&I&I&Z&Z \end{array} \end{array}$} & \gate{X^{\red{s_4}}}\slice[style={blue!70},label style={pos=1, anchor=north}]{$\begin{array}{c}\ket{\psi_2}={}\\ \begin{array}{lcccc} (-1)^{\red{s_1}}&X&X&X&X\\ (-1)^{\red{s_2}}&Z&Z&I&I\\ (-1)^{\red{s_2+s_3}}&I&Z&Z&I\\ (-1)^{\red{s_3+s_4}}&I&I&Z&Z \end{array} \end{array}$} & \targ{} & \qw & \qw & \qw\slice[style={blue!70},label style={pos=1, anchor=north}]{$\begin{array}{c}\ket{\psi_3}={}\\ \begin{array}{lcccc} (-1)^{\red{s_1}}&Z&I&I&I\\ (-1)^{\red{s_2}}&I&Z&I&I\\ (-1)^{\red{s_2+s_3}}&I&I&Z&I\\ (-1)^{\red{s_3+s_4}}&I&I&I&Z \end{array} \end{array}$} & \meter{}\rstick{$m_4$}
    \end{quantikz}};
    \node[below=0mm of a] {\large $m_1 = \red{s_1},\qquad m_2 = \red{s_2},\qquad m_3 = \red{s_2\oplus s_3},\qquad m_4 = \red{s_3\oplus s_4}$};
\end{tikzpicture}}
    \vspace{-2mm}
    \caption{Overview of phase symbolization. Pauli faults in stabilizer circuits only affect the phases of stabilizer generators. As a result, possible
    Pauli faults can be accumulated explicitly in the phases with symbolic expressions, making measurement outcomes into symbolic expressions. With these symbolic expressions, we only need to substitute symbolic variables with concrete values to achieve sampling measurement outcomes, thus avoiding the cost of repeatedly traversing the circuit.}\label{fig:overview}
    \vspace{-6mm}
\end{figure}
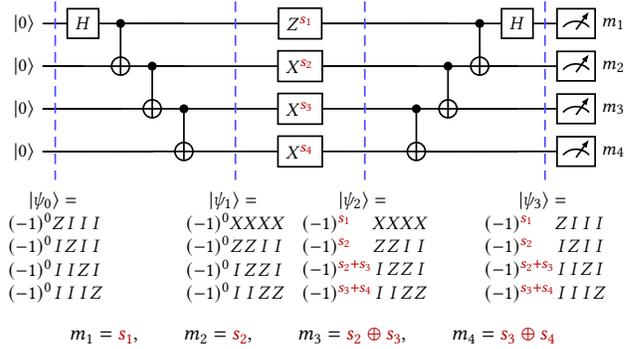

To address the difficulty of generating large samples of measurement outcomes, we propose a novel idea of phase symbolization for simulating stabilizer circuits.
In standard stabilizer circuit simulations~\cite{aaronson2004improved,gottesman1997stabilizer,gidney2021stim}, where evolutions of quantum states are tracked with stabilizer generators (see the lists of Pauli strings in \cref{fig:overview}),
we note that Pauli gates only affect the phase of stabilizer generators.
For example, in \cref{fig:overview}, the gate $Z^{s_1}$ with $s_1\in \{0,1\}$ only changes a phase $(-1)^0$ of $\ket{\psi_1}$ to the phase $(-1)^{s_1}$ of $\ket{\psi_2}$.
As a result, possible Pauli faults in stabilizer circuits can be accumulated in the phase with symbolic variables as shown in \cref{fig:overview}, where $\ket{\psi_1}$ becomes $\ket{\psi_2}$ after passing through $Z^{s_1}, X^{s_2}, X^{s_3}$ and $X^{s_4}$.
The introduction of this symbolization will not change the control flow of the standard stabilizer circuit simulation algorithm, thus we can easily extend existing algorithms with phase symbolization, but it will make the measurement outcomes into some symbolic expressions as these $m_1,m_2,m_3,m_4$ in \cref{fig:overview}.
With these symbolic expressions, we can clearly see how the faults in the circuit affect the measurement outcomes, and we only need to substitute these symbolic variables with concrete values according to the fault model to achieve sampling measurement outcomes, thus avoiding the cost of repeatedly traversing the circuit.

\subsubsection*{Contribution and outline}
After reviewing some background knowledge (\S\ref{sec:background}), our major contributions are presented as follows:
\begin{itemize}
    \item With the phase symbolization, an algorithm (\cref{alg}) for efficient sampling outcomes of stabilizer circuits that traverses the circuit only once is proposed (\S\ref{sec:symbolic_phases}). Specifically, we describe how to integrate symbolic phases into the stabilizer tableau and maintain them efficiently through bit-vector encoding; and turn the sampling process into bit-matrix multiplication.
    \item For efficient implementation of our algorithm and also other stabilizer simulation algorithms based on stabilizer tableau, we 
    propose a new data layout of the stabilizer tableau in the memory (\S\ref{sec:implementation}), which has later been experimentally verified to have advantages over previous tools in some cases.
    \item We implement our algorithm and data layout in a Julia package named \symphase{} and evaluate its ability to surpass the state-of-the-art simulator, Stim, for sampling stabilizer circuits on several benchmarks (\S\ref{sec:evaluation}).
\end{itemize}

\subsubsection*{Related work}
Stabilizer circuit simulation is a well-studied topic in quantum computing.
A key method for simulating stabilizer circuits is the stabilizer tableau method proposed by~\cite{gottesman1998heisenberg} and improved by~\cite{aaronson2004improved}. 
To speed up the sampling of stabilizer circuits with Pauli faults, a technique called Pauli frame was introduced by~\cite{PhysRevA.99.062337}, which tracks the difference between the state with and without faults, and reduces the number of Pauli strings that need to be propagated for sampling an $n$-qubit circuit from $n$ to $1$.
This method was also adopted by Stim, the state-of-the-art stabilizer simulator~\cite{gidney2021stim}.
Recently, \citet{delfosse2023simulation} proposed a method that can extract the relationship between faults and measurement outcomes by traversing the circuit backward once, which greatly improves the sampling efficiency compared to previous work.
Our work achieves the same result by traversing the circuit forward once.
But the basic ideas of~\cite{delfosse2023simulation} and ours are fundamentally different.  
A comparison of the complexity of~\citet{delfosse2023simulation} with ours is presented in \cref{tab:complexity}. 

\section{Background}\label{sec:background}
This paper presupposes some basic knowledge of quantum computing, such as quantum bits (qubits) and quantum circuits.
Readers who are unfamiliar with these concepts can refer to the textbook by~\citet[Chapter 2, 4]{nielsen2010quantum}.

\subsection{Stabilizer Circuits}
\subsubsection*{Pauli strings}
There are four Pauli matrices:
\begin{align*}
    I = \begin{pmatrix}
        1 & 0\\0 & 1
    \end{pmatrix}, \quad
    X = \begin{pmatrix}
        0 & 1\\ 1 & 0
    \end{pmatrix}, \quad
    Y = \begin{pmatrix}
        0 & -i\\ i & 0
    \end{pmatrix}, \quad
    Z = \begin{pmatrix}
        1 & 0\\ 0 & -1
    \end{pmatrix}.
\end{align*}
An $n$-qubit Pauli string is a tensor product of $n$ Pauli matrices with a phase of $\pm 1$ or $\pm i$,
e.g., $-XYZI = -X\otimes Y\otimes Z\otimes I$ is a $4$-qubit Pauli string. We usually omit tensor product signs.
To simplify the notation in dealing with multiple qubits, we also omit the $I$ matrices in Pauli strings and use subscripts to indicate the qubits that the non-identity Pauli matrices act on.
For example, $X_1Y_2Z_3$ means applying $X$ to qubit $1$, $Y$ to qubit $2$, $Z$ to qubit $3$ and $I$ to the rest of qubits;
when restricted to $4$ qubits, $X_1Y_2Z_3$ is regarded as $XYZI$.

\subsubsection*{Stabilizer generators and stabilizer states}
A state $\ket{\psi}$ is \emph{stabilized} by a unitary $U$ if $U\ket{\psi} = \ket{\psi}$, i.e., $\ket{\psi}$ is an eigenvector of $U$ with eigenvalue $1$.
For example, the minus state $\ket{-}$ is stabilized by $-X$ and the bell state $\ket{\beta_{00}} = \frac{1}{\sqrt{2}}(\ket{00}+\ket{11})$ is stabilized by $XX$.
In this paper, we only consider states stabilized by Pauli strings.
For an $n$-qubit state $\ket{\psi}$, let $\stab(\ket{\psi})$ denote the set of all $n$-qubit Pauli strings that stabilize $\ket{\psi}$.
For any $P, Q \in \stab(\ket{\psi})$, we can easily check that $P \cdot Q, P\cdot Q^{-1} \in \stab(\ket{\psi})$, thus $\stab(\ket{\psi})$ is also a group and we call it the \emph{stabilizer group} of $\ket{\psi}$.
The independent generators, which are all Pauli strings, of the stabilizer group are called \emph{stabilizer generators}.

An $n$-qubit state $\ket{\psi}$ is called a \emph{stabilizer state} if $\stab(\ket{\psi})$ has $n$ stabilizer generators.
In this case, with global
phase ignored, $\ket{\psi}$ is the only $n$-qubit state stabilized by $\stab(\ket{\psi})$.
Therefore, there is a one-to-one correspondence between a stabilizer state $\ket{\psi}$ and its stabilizer group $\stab(\ket{\psi})$.

\subsubsection*{Clifford gates and stabilizer circuits}
For a state $\ket{\psi}$ and a Pauli string $P$ that stabilizes $\ket{\psi}$, a unitary $U$ transforms $\ket{\psi}$ to $U\ket{\psi}$, which can be reflected by the transformation from $P$ to $UPU^{\dagger}$ (conjugation by $U$) as $U\ket{\psi}$ is stabilized by $UPU^{\dagger}$.
To ensure that $UPU^{\dagger}$ is still a Pauli string, we consider those unitaries $U$ that conjugate Pauli strings to Pauli strings, i.e., for any Pauli string $P$, $UPU^{\dagger}$ is still a Pauli string.
Such unitaries are called \emph{Clifford gates} and can be constructed from the three gates~\cite{gottesman1997stabilizer}:
\begin{align*}
    H = \frac{1}{\sqrt{2}}\begin{pmatrix}
        1 & 1 \\ 1 & -1
    \end{pmatrix}, \quad
    S = \begin{pmatrix}
        1 & 0 \\ 0 & i
    \end{pmatrix}, \quad
    \cnot = \begin{psmallmatrix}
        1 & 0 & 0 & 0 \\ 0 & 1 & 0 & 0 \\
        0 & 0 & 0 & 1 \\ 0 & 0 & 1 & 0
    \end{psmallmatrix}
\end{align*}

A \emph{stabilizer circuit} is a quantum circuit that uses $H, S, \mathit{CNOT}$ gates (Clifford gates), computational measurements, and $\ket{0}^{\otimes n}$ as the initial state.
The central idea of \emph{stabilizer formalism}~\cite[Chapter 10.5]{nielsen2010quantum} is to describe a state $\ket{\psi}$ by using its stabilizer group $\stab(\ket{\psi})$, which can be identified by stabilizer generators.
For stabilizer circuits, the initial state $\ket{0}^{\otimes n}$ is a stabilizer state with $n$ stabilizer generators $Z_1,Z_2, \ldots, Z_n$;
the Clifford gates and computational measurements will turn it into states that also admit $n$ stabilizer generators~\cite[Chapter 10.5]{nielsen2010quantum}.
This idea provides an efficient simulation of stabilizer circuits by tracking stabilizer generators, sometimes known
as the \emph{Gottesman-Knill theorem}~\cite{gottesman1998heisenberg}.

\subsection{Stabilizer Tableau Simulation}
The most well-known approach to simulating $n$-qubit stabilizer circuits is to maintain an $n\times (2n+1)$ \emph{stabilizer tableau} $\left(
    \bm{X} \mid \bm{Z} \mid \bm{R}
\right)$ that encodes $n$ stabilizer generators $P_1,\ldots,P_n$ as follows.
{\small
\begin{align*}
    \left(
        \bm{X} \mid \bm{Z} \mid \bm{R}
    \right) = 
    \left(\!\setlength{\arraycolsep}{4pt}
    \begin{array}{ccc|ccc|c}
        x_{11} & \cdots & x_{1n} & z_{11} & \cdots & z_{1n} & r_1 \\[-1mm]
        \vdots & \ddots & \vdots & \vdots & \ddots & \vdots & \vdots \\[-1mm]
        x_{n1} & \cdots & x_{nn} & z_{n1} & \cdots & z_{nn} & r_n
    \end{array}
    \!\right)
\end{align*}
}
The $i$-th row of the stabilizer tableau corresponds to a stabilizer generator $P_i$, where the bit-pairs $x_{ij}z_{ij} = 00, 10, 01, 11$ denote the $j$-th Pauli matrix on the $j$-th qubit: $00$ means $I$, $10$ means $X$, $01$ means $Z$ and $11$ means $Y$;
the bit $r_i = 0$ or $1$  for positive or negative phase, respectively.

The updates corresponding to Clifford gates $H$, $S$, $\cnot$ require only $\mathcal{O}(n)$ time.
For example, an $H$ gate on qubit $a$ will set $r_i \coloneq r_i \oplus x_{ia}z_{ia}$ and swap $x_{ia}$ with $z_{ia}$ for all $i\in\{1,\ldots,n\}$, which matches the conjugation by $H$ to Pauli matrices: $HXH^{\dagger} = Z, HZH^{\dagger} = X, HYH^{\dagger} = -Y$.
However, the updates corresponding to computational basis measurements take $\mathcal{O}(n^3)$ time in practice~\cite{aaronson2004improved}, which is in polynomial time but does not scale well enough.

\subsubsection*{The improved tableau algorithm}
To improve the complexity of computational basis measurements in tableau simulation, \citet{aaronson2004improved} (A-G) introduced \emph{destabilizer generators} to stabilizer tableau as follows.
{\small
\begin{equation}\label{eq:stabilizer1}
    \left(
    \begin{array}{c|c|c}
        \bar{\bm{X}} & \bar{\bm{Z}} & \bar{\bm{R}} \\
        \hline
        \bm{X} & \bm{Z} & \bm{R}
    \end{array}
    \right)
    = 
    \left(\!\setlength{\arraycolsep}{4pt}
    \begin{array}{ccc|ccc|c}
        \bar{x}_{11} & \cdots & \bar{x}_{1n} & \bar{z}_{11} & \cdots & \bar{z}_{1n} & \bar{r}_1 \\[-1mm]
        \vdots & \ddots & \vdots & \vdots & \ddots & \vdots & \vdots \\[-1mm]
        \bar{x}_{n1} & \cdots & \bar{x}_{nn} & \bar{z}_{n1} & \cdots & \bar{z}_{nn} & \bar{r}_n \\
        \hline
        x_{11} & \cdots & x_{1n} & z_{11} & \cdots & z_{1n} & r_1 \\[-1mm]
        \vdots & \ddots & \vdots & \vdots & \ddots & \vdots & \vdots \\[-1mm]
        x_{n1} & \cdots & x_{nn} & z_{n1} & \cdots & z_{nn} & r_n
    \end{array}
    \!\right)
\end{equation}
}
The upper half of the tableau $\left(
    \bar{\bm{X}} \mid \bar{\bm{Z}} \mid \bar{\bm{R}}
\right)$ represents $n$ destabilizer generators $\bar{P}_1, \ldots, \bar{P}_n$ such that $\bar{P}_i$ anticommutes with $P_i$ and commutes with $P_j$ for $j\neq i$.
With the help of destabilizer generators, the updates corresponding to computational basis measurements can be realized by a series of row operations (multiply two Pauli strings).
Then the complexity of computational basis measurements is reduced to $\mathcal{O}(n^2)$ time.
\section{Phase Symbolization for Stabilizer Tableau}\label{sec:symbolic_phases}
To speed up the simulation of stabilizer circuits, we introduce the concept of phase symbolization, which is based on the following key facts (\cref{fact:1,fact:2}). 

\begin{fact}\label{fact:1}
    The Pauli gates $X, Y, Z$ \emph{only affect the phase part $\bm{R}, \bar{\bm{R}}$ of the stabilizer tableau}.
    Specifically, for all $i\in\{1,\ldots,n\}$, 
    \begin{itemize}
        \item $X$ gate on qubit $a$: set $r_i\coloneq r_i \oplus z_{ia}, \bar{r}_i\coloneq \bar{r}_i\oplus  \bar{z}_{ia}$;
        \item $Y$ gate on qubit $a$: set $r_i\coloneq r_i\oplus x_{ia}\oplus z_{ia}, \bar{r}_i\coloneq \bar{r}_i\oplus \bar{x}_{ia}\oplus \bar{z}_{ia}$;
        \item $Z$ gate on qubit $a$: set $r_i\coloneq r_i\oplus x_{ia}, \bar{r}_i\coloneq \bar{r}_i\oplus \bar{x}_{ia}$.
    \end{itemize}
\end{fact}

\begin{fact}\label{fact:2}
    The control flow of A-G's algorithm is independent of the values of $\bm{R}, \bar{\bm{R}}$, i.e., all branches in this algorithm are determined by $\bm{X}, \bm{Z}, \bar{\bm{X}}, \bar{\bm{Z}}$ of the tableau.
    The values of $\bm{R}, \bar{\bm{R}}$ can only affect the outcomes of measurements.
\end{fact}

Combining \cref{fact:1,fact:2}, whether a Pauli gate is applied to a qubit will be reflected in whether some rows of $\bm{R}$ and $\bar{\bm{R}}$ are flipped;
hence, it will decide whether the later measurement outcomes are flipped.
This phenomenon is also formalized into the \emph{Pauli frame propagation}~\cite{PhysRevA.99.062337} in Stim~\cite{gidney2021stim}:
To simulate a stabilizer circuit with Pauli faults, we first generate the noiseless measurement outcomes and then use the Pauli frame, a Pauli string that propagates on the circuit, to track the difference between the noiseless state and a sampled noisy state.
This Pauli frame allows us to sample which measurements should be flipped by the noises.
Since tracking the Pauli frame requires maintaining only one Pauli string, the subsequent sampling process takes $\mathcal{O}(1)$ time per gate and measurement.

However, the Pauli frame propagation needs to go through the circuit for each sampling.
In contrast, based on \cref{fact:1,fact:2}, \emph{we can identify which measurements in the circuit are affected by the preceding Pauli faults (and Pauli gates) and may need to be flipped}.
Thus, we propose the symbolic phases to 
capture the flipping relationship.

\subsection{Symbolic Phases}\label{subsec:symbolic_phases}
Instead of assigning specific values to the elements in $\bm{R}$ and $\bar{\bm{R}}$ of stabilizer tableau, we use symbolic expressions to represent them and call them symbolic phases.
The stabilizer tableau becomes:
{\small
\begin{equation}\label{eq:stabilizer2}
    \left(
    \begin{array}{c|c|c}
        \bar{\bm{X}} & \bar{\bm{Z}} & \bar{\bm{R}} \\
        \hline
        \bm{X} & \bm{Z} & \bm{R}
    \end{array}
    \right)
    = 
    \left(\!\setlength{\arraycolsep}{4pt}
    \begin{array}{ccc|ccc|c}
        \bar{x}_{11} & \cdots & \bar{x}_{1n} & \bar{z}_{11} & \cdots & \bar{z}_{1n} & \bar{S}_1 \\[-1mm]
        \vdots & \ddots & \vdots & \vdots & \ddots & \vdots & \vdots \\[-1mm]
        \bar{x}_{n1} & \cdots & \bar{x}_{nn} & \bar{z}_{n1} & \cdots & \bar{z}_{nn} & \bar{S}_n \\
        \hline
        x_{11} & \cdots & x_{1n} & z_{11} & \cdots & z_{1n} & S_1 \\[-1mm]
        \vdots & \ddots & \vdots & \vdots & \ddots & \vdots & \vdots \\[-1mm]
        x_{n1} & \cdots & x_{nn} & z_{n1} & \cdots & z_{nn} & S_n
    \end{array}
    \!\right)
\end{equation}
}
where $S_i, \bar{S}_i$ are symbolic expressions over bit-symbols and bit-values with operator $\oplus$.
For a better understanding, let us 
    consider the following simple example circuit:
    
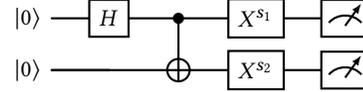
\[\begin{quantikz}[row sep=0.18cm]
        \lstick{\ket{0}} & \gate{H} & \ctrl{1} & \gate{X^{s_1}} & \meter{} \\
        \lstick{\ket{0}} & \qw & \targ{} & \gate{X^{s_2}} & \meter{}
    \end{quantikz}\]
    where $s_1, s_2$ are two bit-symbols indicating whether or not to apply the $X$ gate.
    $X^{s_1}$ and $X^{s_2}$ characterize the possible behaviors of $X$-error on a single qubit.
    By A-G's algorithm, the stabilizer tableau for this circuit evolves as follows:
    \[\begin{gathered}
    \scalebox{0.9}{$\left(\!\setlength{\arraycolsep}{1.5pt}
    \begin{array}{cc|cc|c}
        1 & 0 & 0 & 0 & 0 \\[-4pt]
        0 & 1 & 0 & 0 & 0 \\[-1pt]
        \hline
        0 & 0 & 1 & 0 & 0 \\[-4pt]
        0 & 0 & 0 & 1 & 0
    \end{array}
    \!\right)$} \overto{H_1}
    \scalebox{0.9}{$\left(\!\setlength{\arraycolsep}{1.5pt}
    \begin{array}{cc|cc|c}
        {0} & 0 & {1} & 0 & 0 \\[-4pt]
        0 & 1 & 0 & 0 & 0 \\[-1pt]
        \hline
        {1} & 0 & {0} & 0 & 0 \\[-4pt]
        0 & 0 & 0 & 1 & 0
    \end{array}
    \!\right)$} \overto{\cnot_{1,2}}
    \scalebox{0.9}{$\left(\!\setlength{\arraycolsep}{1.5pt}
    \begin{array}{cc|cc|c}
        0 & 0 & 1 & 0 & 0 \\[-4pt]
        0 & 1 & 0 & 0 & 0 \\[-1pt]
        \hline
        1 & {1} & 0 & 0 & 0 \\[-4pt]
        0 & 0 & {1} & 1 & 0
    \end{array}
    \!\right)$} \overto{X^{s_1}_1}
    \scalebox{0.9}{$\left(\!\setlength{\arraycolsep}{1.5pt}
    \begin{array}{cc|cc|c}
        0 & 0 & 1 & 0 & \raisebox{.5pt}{${s_1}$} \\[-4pt]
        0 & 1 & 0 & 0 & 0 \\[-1pt]
        \hline
        1 & 1 & 0 & 0 & 0 \\[-4pt]
        0 & 0 & 1 & 1 & \raisebox{.5pt}{${s_1}$}
    \end{array}
    \!\right)$} \\
    \setlength{\arraycolsep}{1.5pt}
    \begin{array}{cccc}
        \overto{X^{s_2}_2}
    \scalebox{0.9}{$\left(\!\setlength{\arraycolsep}{1.5pt}
    \begin{array}{cc|cc|c}
        0 & 0 & 1 & 0 & \raisebox{.5pt}{$s_1$} \\[-4pt]
        0 & 1 & 0 & 0 & 0 \\[-1pt]
        \hline
        1 & 1 & 0 & 0 & 0 \\[-4pt]
        0 & 0 & 1 & 1 & \raisebox{.5pt}{$s_1{\oplus s_2}$}
    \end{array}
    \!\right)$} \overto{M_1} &
    \scalebox{0.9}{$\left(\!\setlength{\arraycolsep}{1.5pt}
    \begin{array}{cc|cc|c}
        1 & {1} & {0} & 0 & {0} \\[-4pt]
        0 & 1 & 0 & 0 & 0 \\[-1pt]
        \hline
        0 & 0 & {1} & 0 & \raisebox{.5pt}{${s_3}$} \\[-4pt]
        0 & 0 & 1 & 1 & \raisebox{.5pt}{$s_1\oplus s_2$}
    \end{array}
    \!\right)$} & \overto{M_2} &
    \scalebox{0.9}{$\left(\!\setlength{\arraycolsep}{1.5pt}
    \begin{array}{cc|cc|c}
        1 & 1 & 0 & 0 & 0 \\[-4pt]
        0 & 1 & 0 & 0 & 0 \\[-1pt]
        \hline
        0 & 0 & 1 & 0 & \raisebox{.5pt}{$s_3$} \\[-4pt]
        0 & 0 & 1 & 1 & \raisebox{.5pt}{$s_1\oplus s_2$}
    \end{array}
    \!\right)$} \\
    & m_1 = {s_3} && m_2 = {s_1\oplus s_2\oplus s_3}
    \end{array}
    \end{gathered}\]
    \begin{itemize}
        \item The measurement on the first qubit has random outcomes; thus, we introduce a new bit-symbol $s_3$ to indicate outcome $m_1 = s_3$; this $s_3$ is kept in the stabilizer tableau and used by future operations.
        \item After measuring the first qubit, we can find that the measurement on the second qubit is determined, and it results in an outcome $m_2 = s_1\oplus s_2\oplus s_3$ 
    \end{itemize}
    With the symbolic expressions $m_1 = s_1, m_2 = s_1\oplus s_2\oplus s_3$, we can sample concrete values of $s_1,s_2,s_3$ and substitute them in expressions to obtain samples of measurement outcomes.
    These symbols fall into two categories:
    \begin{itemize}
        \item Symbols induced by random measurements, e.g., the symbol $s_3$ above, are sampled to $0$ and $1$ with probabilities $1/2$ and $1/2$, respectively.
        \item Symbols induced by Pauli faults are sampled specifically according to the probability of the occurrence of Pauli strings in the Pauli faults.
        For example, a single-qubit $X$-error $\cE(\rho) = (1-p)\rho + pX\rho X$ with parameter $p$ corresponds to $X^{s}$ and the bit-symbol $s$ will be sampled to $0$ and $1$ with probabilities $1-p$ and $p$, respectively; a single-qubit depolarization $\cD(\rho) = (1-p)\rho + \frac{p}{3}X\rho X + \frac{p}{3}Y \rho Y + \frac{p}{3}Z\rho Z$ corresponds to $X^{s_1}Z^{s_2}$ and the bit-symbols $s_1s_2$ will be sampled to $00$, $10$, $01$ and $11$ with probabilities $1-p$, $p/3$, $p/3$ and $p/3$, respectively.
    \end{itemize}

For general stabilizer circuits with Pauli faults, the introduction of symbolic phases will turn all the outcomes of measurements into symbolic expressions.
Sampling the measurement outcomes becomes substituting the symbols according to probability and evaluating the symbolic expressions.
This approach avoids the cost of repeatedly traversing the circuit like Pauli frame propagation.

\subsection{Tableau Algorithm with Symbolic Phases}
Now that we have introduced symbolic phases, let us see how to maintain them efficiently during the simulation process and how to speed up the sampling of stabilizer circuits with more details.

\subsubsection{Representing symbolic expressions with bit-vectors}
Since the symbolic expressions here only involve bit-symbols and operator $\oplus$, we can use bit-vectors to represent them.
Considering that the circuit will introduce at most $n_s$ symbols, we represent each bit-symbol $s_j, 1\leq j\leq n_s$, with a bit-vector $\bm{s}_j$:
\[s_j \mapsto \bm{s}_j = \begin{pmatrix}
    \delta_{0,j} & \delta_{1,j} & \cdots & \delta_{n_s,j}
\end{pmatrix} \in \FF_2^{n_s+1},\]
where $\delta_{i,j} = 1$ if $i=j$ and $\delta_{i,j} = 0$ if $i\neq j$.
In particular, we add a symbol $s_0$ to represent the constant $1$.
Then a symbolic expression $S = s_{j_1}\oplus s_{j_2} \oplus \cdots \oplus s_{j_k}$, $0\leq j_1\leq \cdots \leq j_k \leq n_s$, is represented by a bit-vector $\bm{S}$:
\[S \mapsto \bm{S} = \bm{s}_{j_1} + \bm{s}_{j_2} + \cdots + \bm{s}_{j_k} \in \FF_2^{n_s+1},\]
where $+$ is the addition operator in $\FF_2^{n_s+1}$, or it can be referred to the bitwise XOR.
Therefore, the symbolic phases $\bar{S}_j, S_j$ in \cref{eq:stabilizer2} are represented by bit-vectors $\bm{S}_j, \bar{\bm{S}}_j$:
\begin{align*}
    \bar{S}_j & \mapsto \bar{\bm{S}}_j = \begin{pmatrix}
        \bar{s}_{j,0} & \bar{s}_{j,1} & \cdots & \bar{s}_{j, n_s}
    \end{pmatrix} \in \FF_2^{n_s+1}, \\
    S_j & \mapsto \bm{S}_j = \begin{pmatrix}
        s_{j,0} & s_{j,1} & \cdots & s_{j, n_s}
    \end{pmatrix} \in \FF_2^{n_s+1}.
\end{align*}
And each measurement outcome $m_k$, which is also a symbolic expression, is represented by a bit-vector $\bm{m}_k$:
\[m_k \mapsto \bm{m}_k = \begin{pmatrix}
    m_{k,0} & m_{k,1} & \cdots & m_{k, n_s}
\end{pmatrix} \in \FF^{n_s+1}_2.\]

\subsubsection{Extending A-G's algorithm to stabilizer tableau with symbolic phases}
With the above representation, the stabilizer tableau with symbolic phases becomes a $2n\times (2n+n_s+1)$ tableau (bit-matrix):
{\small\begin{align}\label{eq:stabilizer3}
    \left(\!\setlength{\arraycolsep}{3.5pt}
    \begin{array}{ccc|ccc|cccc}
        \bar{x}_{11} & \cdots & \bar{x}_{1n} & \bar{z}_{11} & \cdots & \bar{z}_{1n} & \bar{s}_{1,0} & \bar{s}_{1,1} & \cdots & \bar{s}_{1, n_s} \\[-1mm]
        \vdots & \ddots & \vdots & \vdots & \ddots & \vdots & \vdots & \vdots & \ddots & \vdots \\[-1mm]
        \bar{x}_{n1} & \cdots & \bar{x}_{nn} & \bar{z}_{n1} & \cdots & \bar{z}_{nn} & \bar{s}_{n,0} & \bar{s}_{n,1} & \cdots & \bar{s}_{n, n_s} \\
        \hline
        x_{11} & \cdots & x_{1n} & z_{11} & \cdots & z_{1n} & s_{1,0} & s_{1,1} & \cdots & s_{1, n_s} \\[-1mm]
        \vdots & \ddots & \vdots & \vdots & \ddots & \vdots & \vdots & \vdots & \ddots & \vdots \\[-1mm]
        x_{n1} & \cdots & x_{nn} & z_{n1} & \cdots & z_{nn} & s_{n,0} & s_{n,1} & \cdots & s_{n, n_s}
    \end{array}
    \!\right)
\end{align}}
The first $2n+1$ columns of \cref{eq:stabilizer3} are the same as the original stabilizer tableau (see \cref{eq:stabilizer1}) in A-G's algorithm~\cite{aaronson2004improved}.
We can extend A-G's algorithm to \cref{eq:stabilizer3} as follows.
\begin{itemize}
    \item[]\hspace{-5mm}(\textbf{Init-C}) For Clifford gates, we update the first $2n+1$ columns of \cref{eq:stabilizer3} as A-G's algorithm;
    \item[]\hspace{-5mm}(\textbf{Init-P}) For Pauli faults, we first decompose them into some $X^{s_j}$ and $Z^{s_k}$. Then, for $X^{s_j}$ ($Z^{s_k}$), we treat the first $2n$ columns of \cref{eq:stabilizer3} together with the $j$-th ($k$-th) column as a stabilizer tableau in A-G's algorithm and update it by $X$ ($Z$) gate as A-G's algorithm.
    \item[]\hspace{-5mm}(\textbf{Init-M}) For computational basis measurements, we update the first $2n+1$ columns of \cref{eq:stabilizer3} as A-G's algorithm: when it comes to adding a phase $s_{j,0}$ to another phase $s_{k,0}$ for some $j,k$, we also add the remaining $s_{j,1}, \ldots, s_{j,n_s}$ to $s_{k,1}, \ldots, s_{k,n_s}$, respectively. \begin{itemize}
        \item If the measurement outcome is random, we fix it to $0$ and apply an $X^{s}$ at the measured qubit, where $s$ is a bit-symbol with sampling probabilities of $1/2$ and $1/2$ for $0$ and $1$, respectively. Then, we record the bit-vector $\bm{s}\in\FF_2^{n_s+1}$ for the symbol $s$ as this measurement outcome.
        \item If the measurement outcome is determined, the measurement outcome output by A-G's algorithm is a summation over phases of some rows of stabilizer generators: $s_{j_1, 0}\oplus s_{j_2,0}\oplus \cdots \oplus s_{j_k,0}$. Since we also track addition operations for the remaining $n_s$ elements of each row, we record the bit-vector $\bm{S}_{j_1}+\bm{S}_{j_2}+\cdots+\bm{S}_{j_k} \in \FF_2^{n_s+1}$ as this measurement outcome.
    \end{itemize}
\end{itemize}

\subsubsection{Sampling measurement outcomes as matrix multiplication}
After traversing the circuit by using (\textbf{Init-C}), (\textbf{Init-P}) and (\textbf{Init-M}), we will get an array of bit-vectors $\bm{m}_1, \bm{m_2}, \ldots, \bm{m}_{n_m} \in \FF^{n_2+1}_2$ representing the measurement outcomes.
For all bit-symbols $s_1, \ldots, s_{n_s}$, we sample a bit-vector
$
\bm{b} = \begin{pmatrix}
    b_0 & b_1 & \cdots & b_{n_s}
\end{pmatrix} \in \FF_2^{n_s+1}
$
as mentioned in \cref{subsec:symbolic_phases}, where $b_j, 1\leq j\leq n_s$, is the sampled bit-value for the bit-symbol $s_j$ and the first entry of $b_0=1$ is generated for the constant symbol $s_0$.
Then, the sampled measurement outcome for $\bm{m}_j$ is $\bm{m}_j\bm{b}^{\intercal} = \sum_{k=0}^{n_s}m_{j,k}b_{k} \in \FF_2$
.

Further, if we want to generate $n_{\text{smp}}$ samples of measurements outcomes for $\bm{m}_1, \bm{m_2}, \ldots, \bm{m}_{n_m}$, we can first generate $n_{\text{smp}}$ bit-vectors $\bm{b}_1, \bm{b}_{2}, \ldots, \bm{b}_{n_{\text{smp}}}$, then obtain samples of measurements outcomes as matrix multiplication:
\begin{equation}
    \label{eq:sampling}
    \bm{M}_{\text{samples}} = \begin{pmatrix}
        \bm{m}_1 \\
        \bm{m}_2 \\
        \vdots \\
        \bm{m}_{n_m}
    \end{pmatrix} \cdot 
    \begin{pmatrix}
        \bm{b}_1^{\intercal} & \bm{b}_2^{\intercal} & \cdots & \bm{b}_{n_{\text{smp}}}^{\intercal}
    \end{pmatrix} \in \FF_2^{n_m \times n_{\text{smp}}},
\end{equation}
where the $j$-th column of $\bm{M}_{\text{samples}}$ is the $j$-th sample of measurement outcomes.

\subsubsection{Our algorithm}
Based on the previous discussions in this section, we present the \cref{alg}.
The algorithm takes a noisy stabilizer circuit $C$ and an integer $n_{\text{smp}}$ as inputs, where $n_{\text{smp}}$ is the number of samples for the measurement outcomes.
The distribution of Pauli faults in $C$ is given by $\PP_C$.

{\small
\begin{algorithm}
    \KwIn{A noisy stabilizer circuit $C$, a noise model $\PP_C$ for $C$'s Pauil noises, an integer $n_{\text{smp}}$.}
    \KwOut{$n_{\text{smp}}$ samples of all the measurements in the circuit $C$.}
    \SetKwFunction{Initialization}{Initialization}
    \SetKwFunction{Sampling}{Sampling}
    \nl $\bm{m}_1,\bm{m}_2, \ldots, \bm{m}_{n_m} \leftarrow$ \Initialization{$C$}\;
    \nl $\bm{M}_{\text{samples}} \leftarrow$ \Sampling{$n_{\text{\rm sample}}, \PP_C, \bm{m}_1, \bm{m}_2, \ldots, \bm{m}_{n_m}$}\;
    \nl \Return{$\bm{M}_{\text{\rm samples}}$}\;
    \SetKwProg{myproc}{Procedure}{}{}
    \setcounter{AlgoLine}{0}
    \nl \myproc{\Initialization{$C$}}{
        \nl Traverse circuit $C$ by using (\textbf{Init-C}), (\textbf{Init-P}) and (\textbf{Init-M}) to obtain $\bm{m}_1, \bm{m}_2, \ldots, \bm{m}_{n_m}$\;
        \nl \Return{$\bm{m}_1, \bm{m}_2, \ldots, \bm{m}_{n_m}$}\;
    }
    \setcounter{AlgoLine}{0}
    \nl \myproc{\Sampling{$n_{\text{\rm sample}}, \PP_C, \bm{m}_1, \bm{m}_2, \ldots, \bm{m}_{n_m}$}}{
        \nl Sample $\bm{b}_1, \bm{b}_2, \ldots, \bm{b}_{n_{\text{smp}}}$ from $\PP_C$\;
        \nl $\bm{M}_{\text{samples}} \leftarrow \begin{psmallmatrix}
            \bm{m}_1^{\intercal} & \bm{m}_2^{\intercal} & \cdots & \bm{m}_{n_m}^{\intercal}
        \end{psmallmatrix}^{\intercal} \cdot \begin{psmallmatrix}
            \bm{b}_1^{\intercal} & \bm{b}_2^{\intercal} & \cdots & \bm{b}_{\text{smp}}^{\intercal}
        \end{psmallmatrix}$\;
        \nl \Return{$\bm{M}_{\text{\rm samples}}$}\;
    }
    \caption{Tableau Algorithm with Symbolic Phases.}\label{alg}
\end{algorithm}
}

Consider an $n$-qubit stabilizer circuit $C$ contains $n_g$ single-qubit and two-qubit gates, $n_m$ computational basis measurements, and $n_p$ single-qubit Pauli faults\footnote{All Pauli faults can be decomposed into single-qubit Pauli faults.}.
The cost of \cref{alg} is divided into two parts:
\begin{itemize}
    \item \Initialization: (\textbf{Init-C}) has a cost of $\cO(n)$ for each gate, thus it takes $\cO(nn_g)$ time; (\textbf{Init-P}) has a cost of $\cO(n)$ for each single-qubit Pauli fault, thus it takes $\cO(nn_p)$ time; $n_m$ measurements and $n_p$ single-qubit Pauli faults introduce at most $n_m+n_p+1$ bit-symbols, then the number of columns in \cref{eq:stabilizer3} is at most $2n+n_m+n_p+1$; (\textbf{Init-M}) has a cost of $\cO(n(n+n_m+n+p))$ for each measurement, thus it takes $\cO(nn_m(n+n_m+n_p))$ time. The total cost of \Initialization is $\cO(nn_g+nn_p+nn_m(n+n_m+n_p))$. Since the cost of A-G's algorithm (without Pauli faults) is $\cO(nn_g+n^2n_m)$, we write the cost of \Initialization as $\cO(nn_g+n^2n_m) + \cO(nn_m(n_m+n_p))$.
    \item \Sampling: We only consider the cost of the line $3$ in it\footnote{We do not take account the cost of sampling $\bm{b}_j$ because this cost is related to the noise model and will be \emph{the same for different algorithms}.}. The number of bit-symbols is at most $n_m+n_p+1$, then the cost is lower than the cost of multiplying a bit-matrix of size $n_m\times (n_m+n_p+1)$ by another bit-matrix of size $(n_m+n_p+1)\times n_{\text{smp}}$, which is $\cO(n_{\text{smp}}n_m(n_m+n_p))$.
\end{itemize}

We compare the complexity of \cref{alg} with the algorithm (Pauli frame propagation~\cite{PhysRevA.99.062337}) used by the state-of-the-art simulator Stim~\cite{gidney2021stim} and the recent ABC sim. algorithm~\cite{delfosse2023simulation} in \cref{tab:complexity}.
They differ as follows:
\begin{itemize}
    \item Compared to Stim's, ABC sim.~and our \cref{alg} incur extra costs of $\cO(n_m(n_g+n_p))$ and $\cO(nn_m(n_m+n_p))$, respectively, for the \Initialization. However, 
    \emph{the overhead of \cref{alg} does not depend on the number of gates ($n_g$), while ABC sim.'s contains $n_g$. Thus, our \cref{alg} is favorable when $n_g$ is large}.
    \item For \Sampling, both ABC sim. and \cref{alg} do not depend on $n_g$, thus ABC sim. and \cref{alg} are improvements over  Stim~\cite{gidney2021stim,PhysRevA.99.062337}.
    However, ABC sim. and \cref{alg} have an additional multiplication factor $\cO(n_m+n_p)$ resulting from matrix multiplication; for the case of sparse circuits, i.e., each measurement outcome is related to a small number of Pauli noises, $\begin{psmallmatrix}
        \bm{m}_1^{\intercal} & \bm{m}_2^{\intercal} & \cdots & \bm{m}_{n_m}^{\intercal}
    \end{psmallmatrix}^{\intercal}$ is a column-sparse matrix, then the cost is reduced to $\cO(n_{\text{smp}}n_m)$.
\end{itemize}

\begin{table}
    \small
    \caption{Complexity comparison of various algorithms for simulating stabilizer circuits. \normalfont Our \cref{alg} is advantageous when the circuits have a large number of quantum gates ($n_g$).}\label{tab:complexity}
    \vspace{-4mm}
    \begin{tabular}{|c|c|c|}
        \hline
        Algorithm & \Initialization & \Sampling${}^{\red{\ddagger}}$ \\
        \hline
        Stim's~\cite{gidney2021stim} & $\cO(nn_g+n^2n_m)$ & $\cO(n_{\text{smp}}(\red{n_g}+n_m+n_p))$ \\
        \hline
        ABC sim.~\cite{delfosse2023simulation} & \thead{$\cO(nn_g+n^2n_m)^{\red{\P}}$ \\  $+$ $\cO(n_m(\red{n_g}+n_p))$} & $\cO(n_{\text{smp}}n_m (n_m+n_p))^{\red{*}}$ \\
        \hline
        \cref{alg} & \thead{$\cO(nn_g+n^2n_m)$ \\ $+$ $ \cO(nn_m(n_m+n_p))$} & $\cO(n_{\text{smp}}n_m(n_m+n_p))^{\red{*}}$ \\
        \hline
    \end{tabular}

    \raggedright
    $n$: number of qubits, $n_g$: number of gates, $n_m$: number of measurements, $n_p$: number of single-qubit Pauli noises, $n_{\text{smp}}$: number of samples.

    ${}^\red{*}$: $\cO(n_{\text{smp}}n_m)$ for sparse circuits.

    ${}^\red{\ddagger}$: The cost of sampling noises from $\PP_C$ is not included because it is the same for all algorithms.

    ${}^{\red{\P}}$: ABC sim. obtains the flipping relationship between measurements and Pauli noises and does not obtain the measurement outcomes without noises. Thus, we should include this term for ABC sim.
    \vspace{-3mm}
\end{table}
\section{Data Layout for Implementation}\label{sec:implementation}
Although there are theoretically efficient algorithms (see \cref{tab:complexity}), they still face some practical issues and challenges in implementing them for real applications.
We next discuss the data layout of the stabilizer tableau for implementation.

\begin{figure}
    \centering
    \tikzset{
        uint/.style={rectangle,draw,line width=.5pt,inner sep=2pt,font=\ttfamily\small,align=center}
    }
    \begin{subfigure}[b]{.52\linewidth}
        \centering
        \hspace{-2mm}\begin{tikzpicture}
    \matrix[nodes={rectangle,draw,line width=.5pt,inner sep=2pt,font=\ttfamily\small},column sep=-.5pt,row sep=-.5pt]
    {
        & \node[draw=none]{$1$}; & \node[draw=none]{$2$}; & \node[draw=none]{$\cdots$}; & \node[draw=none]{$\lceil n/32\rceil$};\\
        \node[draw=none]{$1$}; & \node{ UInt-32}; & \node{UInt-32}; & \node[draw=none]{$\cdots$}; & \node{UInt-32};\\
       \node[draw=none]{$2$}; & \node{UInt-32}; & \node{UInt-32}; & \node[draw=none]{$\cdots$}; & \node{UInt-32};\\
       \node[draw=none]{$3$}; & \node{UInt-32}; & \node{UInt-32}; & \node[draw=none]{$\cdots$}; & \node{UInt-32};\\
       \node[draw=none]{$4$}; & \node{UInt-32}; & \node{UInt-32}; & \node[draw=none]{$\cdots$}; & \node{UInt-32};\\[-2mm]
       \node[draw=none]{$\vdots$};&\node[draw=none]{$\vdots$}; & \node[draw=none]{$\vdots$}; & \node[draw=none,]{$\ddots$}; & \node[draw=none]{$\vdots$}; \\
       \node[draw=none]{$n$}; & \node{UInt-32}; & \node{UInt-32}; & \node[draw=none]{$\cdots$}; & \node{UInt-32};\\
    };
\end{tikzpicture}
        \vspace{1mm}
        \caption{\texttt{chp.c}'s data layout. \normalfont \tikz[baseline]{\node[anchor=base,uint]{UInt-32};} is interpreted as a $1\times 32$ bit-matrix.}\label{fig:chp_tab}
        \vspace{3mm}
    \end{subfigure}\hfill
    \begin{subfigure}[b]{.43\linewidth}
        \centering
        \hfill\llap{\begin{tikzpicture}
    \matrix[nodes={rectangle,draw,line width=.5pt,inner sep=2pt,align=center,font=\ttfamily\small},column sep=-.5pt,row sep=-.5pt]
    {
        & \node[draw=none]{$1$}; & \node[draw=none]{$2$}; & \node[draw=none]{$\cdots$}; & \node[draw=none]{$\lceil n/8\rceil$};\\
        \node[draw=none]{\tiny $1$}; & \node{UInt\\ -64}; & \node{UInt\\ -64}; & \node[draw=none]{$\cdots$}; & \node{UInt\\ -64};\\
       \node[draw=none]{$2$}; & \node{UInt\\ -64}; & \node{UInt\\ -64}; & \node[draw=none]{$\cdots$}; & \node{UInt\\ -64};\\[-2mm]
       \node[draw=none]{$\vdots$};&\node[draw=none]{$\vdots$}; & \node[draw=none]{$\vdots$}; & \node[draw=none,]{$\ddots$}; & \node[draw=none]{$\vdots$}; \\
       \node[draw=none]{$\lceil n/8\rceil$}; & \node{UInt\\ -64}; & \node{UInt\\ -64}; & \node[draw=none]{$\cdots$}; & \node{UInt\\ -64};\\
    };
\end{tikzpicture}}
        \caption{Stim's data layout. \normalfont \tikz[baseline]{\node[anchor=base,uint,font=\ttfamily\tiny]{UInt\\ -64};} is interpreted as an $8\times 8$ bit-matrix.}\label{fig:stim_tab}
    \end{subfigure}
    
    \vspace{-1mm}
    
    \begin{subfigure}[t]{.47\linewidth}
        \centering
        \hfill\llap{\begin{tikzpicture}
    \node[rectangle,draw,line width=1pt,inner sep=2pt,minimum width=7.5mm,minimum height=7.5mm,pattern=horizontal lines] (b){};
    \node[below=-1pt of b,overlay] (c){\rotatebox{90}{$=$}};
    \matrix[below=-1pt of c, nodes={rectangle,draw,line width=.5pt,inner sep=2pt,font=\ttfamily\small},column sep=-.5pt,row sep=-.5pt]
    {
        & \node[draw=none]{$1$}; & \node[draw=none]{$2$}; & \node[draw=none]{$\cdots$}; & \node[draw=none]{$64$};\\
        \node[draw=none]{$1$}; & \node{ UInt-8}; & \node{UInt-8}; & \node[draw=none]{$\cdots$}; & \node{UInt-8};\\
       \node[draw=none]{$2$}; & \node{UInt-8}; & \node{UInt-8}; & \node[draw=none]{$\cdots$}; & \node{UInt-8};\\
       \node[draw=none]{$3$}; & \node{UInt-8}; & \node{UInt-8}; & \node[draw=none]{$\cdots$}; & \node{UInt-8};\\
       \node[draw=none]{$4$}; & \node{UInt-8}; & \node{UInt-8}; & \node[draw=none]{$\cdots$}; & \node{UInt-8};\\[-2mm]
       \node[draw=none]{$\vdots$};&\node[draw=none]{$\vdots$}; & \node[draw=none]{$\vdots$}; & \node[draw=none,]{$\ddots$}; & \node[draw=none]{$\vdots$}; \\
       \node[draw=none]{$512$}; & \node{UInt-8}; & \node{UInt-8}; & \node[draw=none]{$\cdots$}; & \node{UInt-8};\\
    };
\end{tikzpicture}}
        \caption{Data layout for $512\times 512$ bit-matrix. \normalfont \normalfont \tikz[baseline]{\node[anchor=base,uint]{UInt-8};} is interpreted as a $1\times 8$ bit-matrix.}\label{fig:inner_tab}
    \end{subfigure}\hfill
    \begin{subfigure}[t]{.47\linewidth}
        \centering
        \hfill\llap{\begin{tikzpicture}
    \matrix[nodes={rectangle,draw,line width=1pt,inner sep=2pt,minimum width=7.5mm,minimum height=7.5mm,align=center,font=\ttfamily\small},column sep=-1pt,row sep=-1pt] (a)
    {
        & \node[draw=none]{$1$}; & \node[draw=none]{$2$}; & \node[draw=none]{$\cdots$}; & \node[draw=none,overlay]{$\lceil n/512\rceil$};\\
        \node[draw=none]{\tiny $1$}; & \node[pattern=horizontal lines]{}; & \node[pattern=horizontal lines]{}; & \node[draw=none]{$\cdots$}; & \node[pattern=horizontal lines]{};\\
       \node[draw=none]{$2$}; & \node[pattern=horizontal lines]{}; & \node[pattern=horizontal lines]{}; & \node[draw=none]{$\cdots$}; & \node[pattern=horizontal lines]{};\\[-2mm]
       \node[draw=none]{$\vdots$};&\node[draw=none]{$\vdots$}; & \node[draw=none]{$\vdots$}; & \node[draw=none,]{$\ddots$}; & \node[draw=none]{$\vdots$}; \\
       \node[draw=none]{$\lceil n/512\rceil$}; & \node[pattern=horizontal lines]{}; & \node[pattern=horizontal lines]{}; & \node[draw=none]{$\cdots$}; & \node[pattern=horizontal lines]{};\\
    };
\end{tikzpicture}}
        \caption{Our data layout. \normalfont \normalfont \tikz[baseline]{\node[anchor=base,uint,pattern=horizontal lines,minimum width=5mm,minimum height=5mm,line width=1pt]{};} is interpreted as a $512\times 512$ bit-matrix.}\label{fig:our_tab}
    \end{subfigure}
    \caption{Data layout for stabilizer tableau.}
    \vspace{-4mm}
\end{figure}
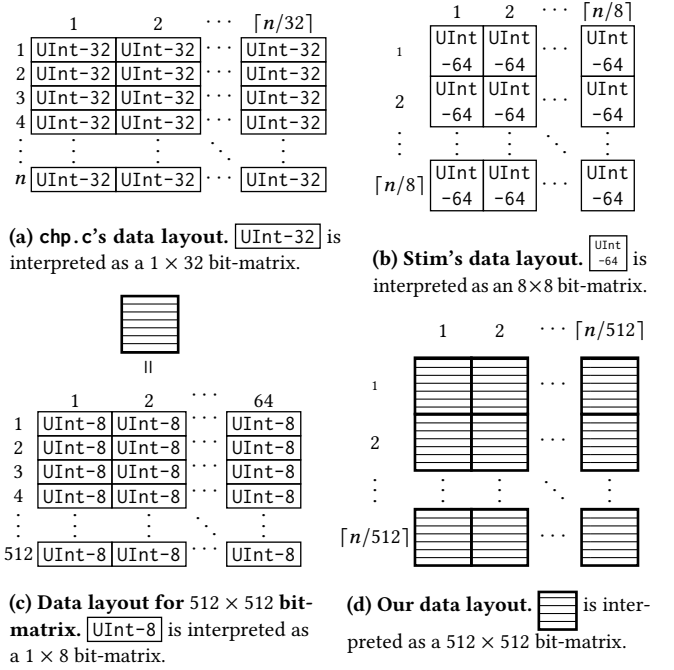

In the implementation \texttt{chp.c}~\cite{chp.c} accompanying with A-G's algorithm~\cite{aaronson2004improved}, the bits were packed into unsigned integers in memory as shown in \cref{fig:chp_tab}, where an \texttt{UInt-32} integer is interpreted as a $1\times 32$ bit-matrix.
Thus, the $n\times \lceil n/32\rceil$ integer-matrix in \cref{fig:chp_tab} can be interpreted as an $n\times n$ bit-matrix.
It offers a compact representation of the stabilizer tableau.
When stored in \emph{row-major order}, it also provides acceleration when we perform row operations for measurements because the rows are contiguous in memory.
However, for quantum gates, which require column operations, the data layout in \cref{fig:chp_tab} is not friendly.

To balance the effects of data layout on row and column operations, Stim~\cite{gidney2021stim} interprets \texttt{UInt-64} integers as $8\times 8$ bit-matrices and places them in column-major order as in \cref{fig:stim_tab}.
This layout with column-major order is friendly to column operations so that quantum gates can be performed quickly.
For measurements, especially a series of measurements, we can transpose it to row-major order temporarily and do a series of measurements before transposing it back for later quantum gates.
Moreover, the contiguous memory in Stim enables the application of SIMD (\emph{Single Instruction, Multiple Data}) operations, which can perform one instruction on multiple data elements (e.g., $256$-bits/$4\times 64$-bits/$8\times 32$-bits) simultaneously.

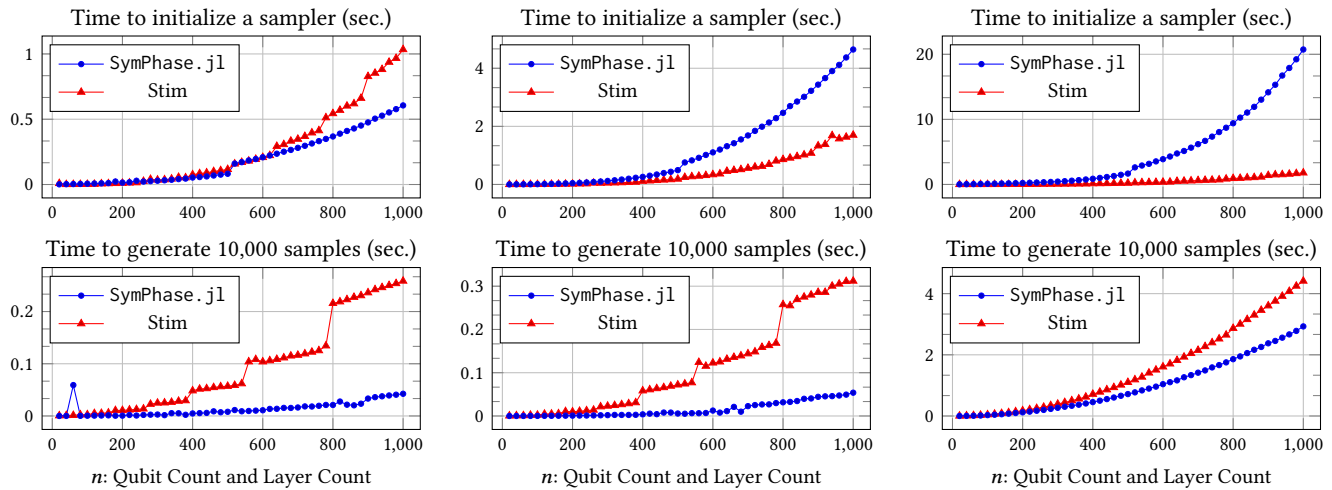
\begin{figure*}[!htbp]
    \begin{subfigure}[t]{.323\linewidth}
        \hfill\llap{\begin{tikzpicture}
    \begin{axis}[
        title={Time to initialize a sampler (sec.)},
        title style={yshift=-7pt},
        reverse legend,
        legend entries={{Stim},{\symphase{}}},
        legend style={font=\small,anchor=west,at={(0.02,0.74)}},
        height=0.65\linewidth,
        width=1.15\linewidth,
        x label style={font=\small},
        x tick label style={font=\footnotesize},
        y tick label style={font=\footnotesize},
        minor y tick num=2,
        enlarge x limits=0.05,
    ]
        \addplot+[
            x filter/.expression={(mod(x, 20) == 0 ? \pgfmathresult : NaN)},
            red,mark=triangle*,mark options={scale=.8,draw=red,fill=red!80!black}
        ] table[x=nq,y=init_time] {./figures/benchmark_stim.dat};
        \addplot+[
            x filter/.expression={(mod(x, 20) == 0 ? \pgfmathresult : NaN)},
            blue,mark=*,mark options={scale=.5,draw=blue,fill=blue!80!black}
        ] table[x=nq,y=init_time] {./figures/benchmark_symphase.dat};
    \end{axis}
\end{tikzpicture}}
    
        \hfill\llap{\begin{tikzpicture}
    \begin{axis}[
        title={Time to generate 10,000 samples (sec.)},
        title style={yshift=-7pt},
        xlabel={$n$: Qubit Count and Layer Count},
        reverse legend,
        legend entries={{Stim},{\symphase{}}},
        legend style={font=\small,anchor=west,at={(0.02,0.74)}},
        height=0.65\linewidth,
        width=1.15\linewidth,
        x label style={font=\small},
        x tick label style={font=\footnotesize},
        y tick label style={font=\footnotesize},
        minor y tick num=2,
        enlarge x limits=0.05,
    ]
        \addplot+[
            x filter/.expression={(mod(x, 20) == 0 ? \pgfmathresult : NaN)},
            red,mark=triangle*,mark options={scale=.8,draw=red,fill=red!80!black}
        ] table[x=nq,y=sample_time] {./figures/benchmark_stim.dat};
        \addplot+[
            x filter/.expression={(mod(x, 20) == 0 ? \pgfmathresult : NaN)},
            blue,mark=*,mark options={scale=.5,draw=blue,fill=blue!80!black}
        ] table[x=nq,y=sample_time] {./figures/benchmark_symphase.dat};
    \end{axis}
\end{tikzpicture}}
        \caption{\normalfont Each layer randomly selects 5 pairs (1 pair in Stim's benchmark) of qubits to apply $\cnot$ gates.}\label{fig:random_circuits1}
    \end{subfigure}
    \hfill
    \begin{subfigure}[t]{.323\linewidth}
        \hfill\llap{\begin{tikzpicture}
    \begin{axis}[
        title={Time to initialize a sampler (sec.)},
        title style={yshift=-7pt},
        reverse legend,
        legend entries={{Stim},{\symphase{}}},
        legend style={font=\small,anchor=west,at={(0.02,0.74)}},
        height=0.65\linewidth,
        width=1.15\linewidth,
        x label style={font=\small},
        x tick label style={font=\footnotesize},
        y tick label style={font=\footnotesize},
        minor y tick num=2,
        enlarge x limits=0.05,
    ]
        \addplot+[
            x filter/.expression={(mod(x, 20) == 0 ? \pgfmathresult : NaN)},
            red,mark=triangle*,mark options={scale=.8,draw=red,fill=red!80!black}
        ] table[x=nq,y=init_time] {./figures/benchmark_stimCNOT.dat};
        \addplot+[
            x filter/.expression={(mod(x, 20) == 0 ? \pgfmathresult : NaN)},
            blue,mark=*,mark options={scale=.5,draw=blue,fill=blue!80!black}
        ] table[x=nq,y=init_time] {./figures/benchmark_symphaseCNOT.dat};
    \end{axis}
\end{tikzpicture}}
    
        \hfill\llap{\begin{tikzpicture}
    \begin{axis}[
        title={Time to generate 10,000 samples (sec.)},
        title style={yshift=-7pt},
        xlabel={$n$: Qubit Count and Layer Count},
        reverse legend,
        legend entries={{Stim},{\symphase{}}},
        legend style={font=\small,anchor=west,at={(0.02,0.74)}},
        height=0.65\linewidth,
        width=1.15\linewidth,
        x label style={font=\small},
        x tick label style={font=\footnotesize},
        y tick label style={font=\footnotesize},
        minor y tick num=2,
        enlarge x limits=0.05,
    ]
        \addplot+[
            x filter/.expression={(mod(x, 20) == 0 ? \pgfmathresult : NaN)},
            red,mark=triangle*,mark options={scale=.8,draw=red,fill=red!80!black}
        ] table[x=nq,y=sample_time] {./figures/benchmark_stimCNOT.dat};
        \addplot+[
            x filter/.expression={(mod(x, 20) == 0 ? \pgfmathresult : NaN)},
            blue,mark=*,mark options={scale=.5,draw=blue,fill=blue!80!black}
        ] table[x=nq,y=sample_time] {./figures/benchmark_symphaseCNOT.dat};
    \end{axis}
\end{tikzpicture}}
        \caption{\normalfont Each layer randomly selects $\lfloor\frac{n}{2}\rfloor$ pairs of qubits to apply $\cnot$ gates.}\label{fig:random_circuits2}
    \end{subfigure}
    \hfill
    \begin{subfigure}[t]{.323\linewidth}
        \hfill\llap{\begin{tikzpicture}
    \begin{axis}[
        title={Time to initialize a sampler (sec.)},
        title style={yshift=-7pt},
        reverse legend,
        legend entries={{Stim},{\symphase{}}},
        legend style={font=\small,anchor=west,at={(0.02,0.74)}},
        height=0.65\linewidth,
        width=1.15\linewidth,
        x label style={font=\small},
        x tick label style={font=\footnotesize},
        y tick label style={font=\footnotesize},
        minor y tick num=2,
        enlarge x limits=0.05,
    ]
        \addplot+[
            x filter/.expression={(mod(x, 20) == 0 ? \pgfmathresult : NaN)},
            red,mark=triangle*,mark options={scale=.8,draw=red,fill=red!80!black}
        ] table[x=nq,y=init_time] {./figures/benchmark_stimCNOT_dep.dat};
        \addplot+[
            x filter/.expression={(mod(x, 20) == 0 ? \pgfmathresult : NaN)},
            blue,mark=*,mark options={scale=.5,draw=blue,fill=blue!80!black}
        ] table[x=nq,y=init_time] {./figures/benchmark_symphaseCNOT_dep.dat};
    \end{axis}
\end{tikzpicture}}
    
        \hfill\llap{\begin{tikzpicture}
    \begin{axis}[
        title={Time to generate 10,000 samples (sec.)},
        title style={yshift=-7pt},
        xlabel={$n$: Qubit Count and Layer Count},
        reverse legend,
        legend entries={{Stim},{\symphase{}}},
        legend style={font=\small,anchor=west,at={(0.02,0.74)}},
        height=0.65\linewidth,
        width=1.15\linewidth,
        x label style={font=\small},
        x tick label style={font=\footnotesize},
        y tick label style={font=\footnotesize},
        minor y tick num=2,
        enlarge x limits=0.05,
    ]
        \addplot+[
            x filter/.expression={(mod(x, 20) == 0 ? \pgfmathresult : NaN)},
            red,mark=triangle*,mark options={scale=.8,draw=red,fill=red!80!black}
        ] table[x=nq,y=sample_time] {./figures/benchmark_stimCNOT_dep.dat};
        \addplot+[
            x filter/.expression={(mod(x, 20) == 0 ? \pgfmathresult : NaN)},
            blue,mark=*,mark options={scale=.5,draw=blue,fill=blue!80!black}
        ] table[x=nq,y=sample_time] {./figures/benchmark_symphaseCNOT_dep.dat};
    \end{axis}
\end{tikzpicture}}
        \caption{\normalfont Each layer randomly selects $\lfloor\frac{n}{2}\rfloor$ pairs of qubits to apply $\cnot$ gates and additionally applies single-qubit depolarize noise to each qubit.}\label{fig:random_circuits3}
    \end{subfigure}
    \vspace{-3mm}
    \caption{Performance results of sampling layered random interaction circuits. \normalfont Each circuit is made up of $n$ qubits with $n$ layers. Each layer randomly applies an $H$, $S$ and $I$ gate to each qubit, then applies $\cnot$ gates, then samples $5\%$ of the qubits to measure in the computational basis. At the end of the circuit, each qubit is measured in the computational basis. The maximum size of the circuit in the experiment reaches 1,000 qubits, 1160,000 quantum gates, 2000,000 Pauli faults, and 51,000 measurements.}\label{fig:random_circuits}
\end{figure*}

\subsubsection*{Our data layout}
Despite the high performance of Stim with its data layout (\cref{fig:stim_tab}), we observe that the transpose operation was time-consuming.
Since the number of measurements in the circuits is usually smaller than the number of gates, we adopt a new layout in \cref{fig:our_tab}.
Each shaded block contains an \texttt{UInt-8} matrix of size $512\times 64$ in column-major order, which represents a $512\times 512$ bit-matrix.
This layout allows SIMD operations for doing column operations (gates).

For row operations (measurements), we only do local transpositions of shaded blocks (\cref{fig:inner_tab}).
Such local transpositions reduce the time required to transpose the entire bit-matrix.
With local transpositions, \cref{fig:inner_tab} is in row-major order.
For the entire bit-matrix, each row is not allocated continuously in memory but separated into groups of $512$ bits.
Although it prevents us from manipulating rows consecutively, the fixed length of $512$ bits already provides sufficient speedup.
\section{Evaluations}\label{sec:evaluation}
We have developed a Julia~\cite{bezanson2017julia} package named \symphase{}\footnote{see \url{https://github.com/njuwfang/SymPhase.jl}.} that implements \cref{alg}. 
\symphase{} uses the data layout shown in \cref{fig:our_tab} for the \Initialization and the sparse implementation of matrix multiplication for the \Sampling.
To demonstrate the efficiency of our \cref{alg} in sampling results of stabilizer circuits and to evaluate the performance of \symphase{}, we chose to compare it with the state-of-the-art stabilizer simulator.

\subsubsection*{\normalfont \bfseries Baseline}
The state-of-the-art stabilizer simulator known to us is Stim~\cite{gidney2021stim}, which has not only surpassed popular simulators such as Qiskit's stabilizer method~\cite{aleksandrowicz2019qiskit}, Cirq's Clifford simulator~\cite{omole2020cirq}, \citet{aaronson2004improved}'s \texttt{chp.c} and GraphSim~\cite{anders2006fast} in performance, but is also being actively developed\footnote{See \url{https://github.com/quantumlib/Stim}.}.

\subsubsection*{\normalfont \bfseries Benchmark}
We selected three classes of randomly generated circuits for the benchmark, which are variants of the benchmark used in Stim~\cite{gidney2021stim}.
This way, we can avoid the influence of circuit structures on the comparison results.
For example, circuits for LDPC codes~\cite{PRXQuantum.2.040101} are sparse, which gives us an advantage.
The detailed descriptions of these circuits are given in the captions of \cref{fig:random_circuits,fig:random_circuits1,fig:random_circuits2,fig:random_circuits3}.

\subsubsection*{\normalfont \bfseries Environment}
All our experiments are carried out on a desktop with Intel(R) Core(TM) i7-9700 CPU@3.00GHz and 16G of RAM, running Ubuntu 22.04.2 LTS.
The version of Stim is 1.12.0 (the latest stable version).

\subsubsection*{\normalfont \bfseries Result}
The experimental results are shown in~\cref{fig:random_circuits1,fig:random_circuits2,fig:random_circuits3}.
We report the time for Stim and \symphase{} to initialize a sampler (i.e., the time to analyze the input circuit and create a sampler for generating the measurement results) and the time for Stim’s and \symphase{}'s samplers to generate 10,000 samples of measurement results.
\symphase{} outperforms Stim in all benchmarks in terms of the sampling time, which validates the advantages of our algorithm (see \cref{alg}) and our package (\symphase{}) for sampling stabilizer circuits.
On the other hand, our algorithm has an overhead for symbolic phases, which makes \symphase{} consume more time than Stim in initializing samplers.
However, this overhead is one-time, and the performance of the sampler is crucial for generating a large number of samples for further analysis.
Moreover, we also observe that in \cref{fig:random_circuits1}, \symphase{} has a better initialization time than Stim, which indicates that our data layout has benefits in certain situations.
This is worth further investigation.
\section{Conclusion}\label{sec:conclusion} 
We have presented phase symbolization for fast simulation of stabilizer circuits without traversing the circuit repeatedly.
With a new layout of the stabilizer tableau, a package \symphase{} is implemented,  which has been experimentally evaluated that it surpasses the existing state-of-the-art tool in sampling stabilizer circuits.
We believe that our techniques can provide a useful tool for simulating and analyzing stabilizer circuits, especially for fault-tolerant quantum computing.

We also expect that our ideas and techniques  can be used in other tools for similar or related tasks. 
Under phase symbolization, the measurement outcomes are some symbolic expressions $e$, which can be used to conditionally apply Pauli gates $X^{e}, Y^{e}, Z^{e}$ as we have done for Pauli faults.
This will allow us to achieve better results in dynamic/sequential stabilizer circuit simulations.
The data layout of stabilizer tableau in memory is crucial for implementation;
our layout and Stim's~\cite{gidney2021stim} layout have different advantages in different circuits, so dynamically determining the layout based on the type/pattern of the circuit will be very helpful to improve the performance of the tool.

\begin{acks}
  We thank Kean Chen for insightful discussions and Craig Gidney for pointing out our previous inappropriate use of Stim.
  This work was partly supported by the \grantsponsor{}{National Natural Science Foundation of China}{} under Grant No.~\grantnum{}{61832015}.
\end{acks}

{\small
\balance
\bibliographystyle{ACM-Reference-Format}
\bibliography{main}
}

\end{document}